\documentclass[11pt,twocolumn,psfig]{article}
\begin{document}
\input{psfig}
%
\advance\oddsidemargin  by -0.5in 
  \advance\evensidemargin by -1in
\marginparwidth          1.9cm
\marginparsep            0.4cm
\marginparpush           0.4cm
\topmargin               -0.2cm
\textheight             21.5cm
\hoffset +15mm
\newcommand{\be}{\begin{equation}}
\newcommand{\ee}{\end{equation}}
\newcommand{\bq}{\begin{eqnarray}}
\newcommand{\eq}{\end{eqnarray}}
\newcommand{\bibit}{\nineit}
\newcommand{\bibbf}{\ninebf}
\def\id{{\rm 1\kern-.21em 1}}
\font\nineit=cmti9
\font\ninebf=cmbx9
\title{\bf {Hawking radiation from Trojan states in muonic Hydrogen in strong laser field} }
\author{M. Kalinski
\\
\\
{\it Department of Chemistry and Biochemistry, Utah State University, Logan, UT 84322}}

\begin{twocolumn}
\maketitle
\noindent
{\bf Abstract} --
We show that the Unruh-Davis effect is measurable from Trojan wavepackets
in muonic Hydrogen as the acceleration on the first muonic Bohr orbit reaches 
$10^{25}$ of the earth acceleration.
It is the biggest acceleration achievable in the laboratory environment which 
have been ever predicted for the cyclotronic configuration.
We calculate the ratio between the  power  of Larmor radiation and the power
of Hawking radiation. The Hawking radiation is measurable even for Rydberg 
quantum numbers of the muon due to suppression of spontaneous emission in
Trojan Hydrogen.
\medskip


\noindent
PACS: 42.50.Hz, 36.10.Dr, 04.70.Dy, 03.65.Ge \\

\noindent
{\bf e-mail} address: mkal@cc.usu.edu


\bigskip
\centerline{1. INTRODUCTION}
\medskip

The possibility to observe the Unruh-Davis effect using single 
electron DeWitt detectors has been discussed quite long time ago \cite{bell,peres,jurgen,boyer}.
The  measurements of depolarization of electrons in storage 
ring have been classified confirming of spin states heat up by 
the field vacuum fluctuation \cite{bell1}
The perturbation of  Thomas precession was shown to be weak as the
time of the full deexcitation of the spin state turned out to exceed the age of the  Universe. 

However very recently microscopic cyclotrons with optical cyclotronic frequency in circularly polarized electromagnetic field have been predicted theoretically and confirmed experimentally in single atoms when well confined electron on circular orbit in moving almost eternally without dispersion in strong circularly polarized (CP) electromagnetic field \cite{tps1,tps}.
It has been also suggested recently that the Unruh-Davis effect can be observed
in radiation of driven electrons in ultra strong laser field \cite{strong}.
The laser field strength necessary for the successful experiment  was proposed to be of the order of $10 TeV/cm$. 
The collaborative effect of Coulomb fields is known to 
cause enhanced ionization and the Coulomb ignition of atomic Van der Waals 
clusters \cite{clust}.
Nuclear fusion was observed due to strain Coulomb field acceleration of energetic  ions from atomic cluster \cite{ditmire}.
The laser dissociation of muonic molecules have been recently also  suggested \cite{corkum}.
In this paper we suggest an alternative  method of observing Hawking 
radiation in strong laser field. In contrast to ultra-strong laser method it is mainly the Coulomb field which provides the gigantic acceleration of the Davis detector and much weaker laser field  prepares the quantum particle to be classical. It is also the centrifugal acceleration which 
extends the time of observation practically to infinity on the scale
of physical times involved.
One may notice that for classical (Trojan) nonrelativistic  Coulomb problem of muonic Hydrogen  the centrifugal acceleration of classical electron on first Bohr orbit  is 
\begin{eqnarray}
a={1 \over 4 \pi \epsilon_0} {e^2 \over \mu  a_{\mu}^2} = 1.90 \times 10^{24} g
\end{eqnarray}
where $a_{\mu}$ is muonic Bohr radium $a_{\mu}= a_0 \mu /m_e$
respectively the estimated Davies temperature on $n$-th Rydbeg orbit 
is
\begin{eqnarray}
T={{\hbar a_n }\over {2 \pi  k c} }= 75830 K /n^4
\end{eqnarray}
where $n$ is the muonic Rydberg quantum number and $a_n$ is  
centrifugal acceleration on $n$-the Rydberg orbit.
It is $1000K$ for $n=3$ and therefore should be easily observable.

\bigskip
\centerline{2. TROJAN WAVEPACKETS}
\medskip

We start our detailed analysis from the harmonic  theory of Trojan wavepackets in muonic Hydrogen \cite{tps1,ebe}.
We neglect the spin for simplification.
Note that even through the lifetime of the muon is $2.1970 \times 10^{-6} s $ the packet can make thousands of revolution for the laser optical frequency, more then the lifetime due to its electromagnetic resonant nature.

Since the ratio between the electron velocity on the first Bohr orbit to 
the speed of light is always $\alpha=1/137$ for arbitrary mass of quantum particle the system still can be described
by the Schr{\"o}dinger equation.
The Schr{\"o}dinger equation for the Trojan wavepacket in circularly polarized field  in the Coulomb field is 
\begin{eqnarray}
({{\bf p}^2 \over {2 \mu}} - {1 \over r} - e x {\cal E}_f+ \omega L_z)\psi= i \hbar {\dot \psi}
\end{eqnarray}
where $\mu$ is the muon mass $\mu=206.7683 m_e$.
The standard stability analysis \cite{tps1,ebe} of the trajectory in rotating frame extended to relativistic case gives the following harmonic Gaussian  wavefunction
\begin{eqnarray}
\psi_0=N e^{I \mu \omega x_0 y/\hbar} e^{-i E_0 t/\hbar} \\
\times e^{-\mu \omega [A (x-x_0)^2 + B y^2 + 2 i C (x-x_0) + D z^2]/2 \hbar} \nonumber
\end{eqnarray} 
with parameters $A$, $B$, $C$, $D$.
The stationary factor is the eigenfunction of the harmonic Hamiltonian
\begin{eqnarray}
H_{h}&=&\hbar \omega_{+} a_{+}^{\dagger} a_{+} - \hbar {\omega_{-}} a_{-}^{\dagger} a_{-} + \hbar \omega_z {a_{z}^{\dagger}} a_{z} \\ \nonumber
&+& const
\end{eqnarray}
where \cite{ebe}
\begin{eqnarray}
\omega_{\pm}=\omega \sqrt{2-q \pm {9 q^2 -8 q}} / \sqrt{2} \\
\omega_z= \omega \sqrt{q} \nonumber
\end{eqnarray}
\begin{eqnarray}
q=e^2/4 \pi \epsilon_0 \mu \omega^2 x_0^3
\end{eqnarray}
end the parameter $x_0$ is the classical center of the wavepacket give implicitly
\begin{eqnarray}
-e^2/4 \pi \epsilon_0 x_0^2 + e {\cal E}_f + \mu \omega^2 x_0 = 0
\end{eqnarray}
Note that it is not exactly  the Hamiltoniam of three harmonic oscillators because of negative sign of $a_{-}^{\dagger}a_{-}$ but 
rather Hamiltonian predicting vacuum collapse if coupled to electromagnetic 
field.
The parameters $A, B$ and $C$ are respectively functions of the
dimensionless parameter $q$ only 
\cite{ebe,ibbsp}
\begin{eqnarray}
A(q)&=&\sqrt{(1+2 q)[4 f(q) - 9q^2]}/3q   \\
B(q)&=&\sqrt{(1-q)[4 f(q) - 9 q^2]} /3q \nonumber \\
C(q)&=&f(q)/3q,   \nonumber  \\
D(q)&=&\sqrt{q} \nonumber
\end{eqnarray}
where
\begin{eqnarray}
f(q)=2+q-2 \sqrt{(1-q)(1+2 q)}
\end{eqnarray}

\bigskip
\centerline{3. HAWKING RADIATION}
\medskip

The probability of the Hawking photon emission by the non-relativistic Trojan wavepacket 
can be calculated similarly to the spontaneous emission rate \cite{ibbsp,terashima} as 
\begin{eqnarray}
&&\Gamma_{UD}= \\
&&{e^2 \over \hbar^2} \sum_{i,j} <\psi_0|x_i|\psi_1><\psi_1|x_j|\psi_0> \nonumber \\ 
&&\times G_{ij}[{(E_1-E_0) \over \hbar}] \nonumber
\end{eqnarray}
where $G_{i,j}(\omega)$ is the Fourier transform field-field correlation function for the accelerated
observer on the circular orbit
\begin{eqnarray}
G_{ij}(\Omega)=\int_{-\infty}^{\infty} e^{- i \omega \tilde \tau} G_{i,j}(\tilde \tau)
\end{eqnarray}
\and
\begin{eqnarray}
G_{ij}(\tau)&=& \\ \nonumber
G_{ij}({\tau}_1 - {\tau_2}) &=& <0|E_i({\tau_1}) E_j({\tau_2})0>
\end{eqnarray}
The Wightman function (electromagnetic field tensor - tensor correlation function of the electromagnetic field in the Minkowski vacuum is 
\begin{eqnarray}
&&<0|F_{\mu \nu}(x) F_{\rho \sigma} (x')|0> \\
&=& 4 {{\hbar c }\over \epsilon_0 \pi} \{ (x-x')^{-6} 
\{(g_{\mu \rho} g_{\nu \sigma} - g_{\mu \sigma} g{ \nu \rho)} \nonumber \\
&-&2 [(x-x')_{\mu} (x-x')_{\sigma} g_{\nu \sigma} \nonumber \\
&-& (x-x')_{\nu} (x-x')_{\rho} g_{\mu \sigma} \nonumber \\
&-& (x-x')_{\mu} (x-x')_{\sigma} g_{\nu \rho} \nonumber \\
&+& (x-x')_{\nu} (x-x')_{\rho} g_{\mu \rho} ] \} \nonumber
\end{eqnarray}
The DeWitt detector response is due to delay of the proper time on the
noninertial trajectory.
The space-time trajectory parameterized by the proper time $\tau$ is for Trojan atom
\begin{eqnarray}
x=(c \tau \gamma, R \cos(\omega \tau \gamma), R \sin(\omega \tau \gamma), 0) \nonumber
\end{eqnarray}
where 
\begin{eqnarray}
\beta&=& {v \over  c} \\
\gamma&=&{1 \over \sqrt{(1- \beta^2)}}
\end{eqnarray}
The field-field correlation function for the observer moving 
on the circular trajectory can be estimated \cite{expl} $(\omega \tau \ll 1)$ as 
\begin{eqnarray}
G_{ij}(\tilde \tau - \tilde \tau^{\prime} )= { 1 \over \epsilon_0} {{4 \hbar c} \over \pi} {1 \over{ c^4 {\gamma}^4}} { 1\over {(\tilde \tau} - {\tilde \tau}^{\prime} )^4 } \delta_{ij}
\end{eqnarray}
The only nonvanishing matrix elements between the first deexcited Trojan state
and the Trojan state are those of $x$ and $y$ coordinates \cite{tps1} and can be expressed by their fluctuations.
Using the result of \cite{ibbsp} one gets
\begin{eqnarray}
<\psi_1|x|\psi_0>&=& \alpha \lambda <\psi_0 |x^2| \psi_0> \nonumber \\
&=&{\alpha \lambda \hbar \over {2 A m \omega}} \nonumber  \\
<\psi_1|y|\psi_0>&=& \alpha <\psi_0 |y^2| \psi_0> \nonumber ]\\
&=& {\alpha \hbar \over {2 B m \omega}}
\end{eqnarray}
where \cite{ibbsp}
\begin{eqnarray}
\alpha&=&  {\sqrt{2 \mu \omega/ \hbar} \over  \sqrt{ \lambda^2/A + 1/B}}= {\tilde \alpha} \sqrt{{2 \mu \omega} \over \hbar}\\
\lambda&=&(1+C)/(A+ \omega_{-}/ \omega).
\end{eqnarray}
The energy difference between the Trojan and first deexcited Trojan state
is
\begin{eqnarray}
(E_1-E_0)= \hbar \omega_{-}
\end{eqnarray}
The probability of emission of Hawking photon (the decay rate of muonic Trojan wavepacket due to the Unruh-Davies effect) is therefore
\begin{eqnarray}
\Gamma_{UD}={1 \over 3}\sqrt{\pi \over 2} {{e^2 \omega^2} \over {\epsilon_0 \mu c^3}} {\tilde \alpha}^2( {\lambda^2 \over A^2} + {1 \over B^2}) {\theta}^3
\end{eqnarray}
with the dimensionless scaling function
\begin{eqnarray}
\theta(q)= \sqrt{2-q - \sqrt{(9q^2-8q)}}/\sqrt{2}=\omega_{-}/\omega
\end{eqnarray}

In order the effect to be observable it must be not much less
the decay rate of the emission which is qualitatively the
same as due to the Larmor radiation of classical charged particle on the
circular orbit \cite{ibbsp}. 

The decay rate of the spontaneous emission of the nonrelativistic Trojan 
wavepacket is given by \cite{ibbsp}
\begin{eqnarray}
\Gamma_{SP}= {e^2 \over { 4 \pi \epsilon_0 \mu c^3}} {\omega^2 \over 3} {{(\mu / A - 1/B)^2 }\over {(\mu^2/A + 1/B)}}(1 +\theta)^3
\end{eqnarray}

Fig.1 shows both the Hawking decay rate and the spontaneous emission 
rate as functions of the scaled electric field ${\cal E} = {\cal E}_f \omega^{-4/3}=(1-q)/q^{1/3}$.
The ratio between $\Gamma_{UD}/\Gamma_{SP}$ is therefore
dimensionless parameter depending only on parameter
$q$ of the harmonic theory of the Trojan wavepacket.
The Davis effect is observable when power of spontaneous (Larmor) radiation  is comparable with the  power of Hawking radiation.
Note that the relativistic parameter $\gamma$ from the relativistic Wightman function  within the theory limit  presented
here should  be approximated by  
\begin{figure}
\centerline{\psfig{figure=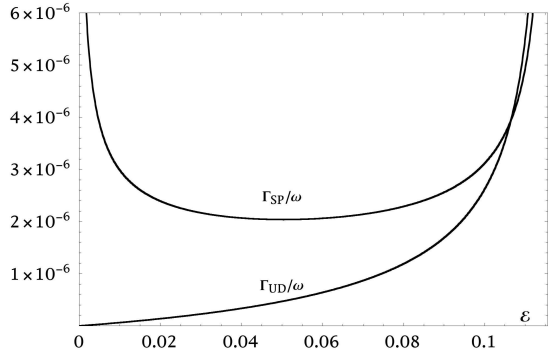,height=4.5in,rheight=3.2in,clip=t}}
\caption{ The relative spontaneous emission rate (for the reference only, see \cite{ibbsp})  and Hawking emission rate due to Unruh-Davies  effect 
as functions of scaled electric field for hypothetical $n=1$.
Both rates scale like $1/n^3$ with the resonant Rydberg number $n$.
Note that singularities \cite{note1} at
stability border values are nonphysical but the Hawking decay rate goes to zero
when there is no electric field present.} 
\end{figure}
$1$ for consistency  ($\beta=1/137)$ since we neither
use the Klein-Gordon or Dirac equation for the theory of the Trojan electron itself.
Fig.2 shows the ratio between the Hawking decay rate and the spontaneous emission rate as function of the scaled electric field $\cal E$.
For the best confined wavepacket we get $\Gamma_{UD}=1.11784 \times 10^6 s^{-1}$
already for $n=12$, almost twice faster then the decay of the muon itself, well sufficient to observe decay of the wavepacket before the decay of the muon and still during thousands of the packet revolutions. It corresponds to the laser intensity $I=5.89 \times 10^{14} W/cm^2$
and the wavelength $\lambda=380.782nm$
Since the  spontaneous emission is the Larmor radiation in the classical limit \cite{ibbsp} the Unruh-Davies emission should be readily observable as the correction to the only  mechanism of radiation.
Note that the Unruh-Davies radiation vanishes in the limit of vanishing 
electric CP field which means that stationary circular states do not emit
Hawking radiation at all as they are not localized DeWitt detectors.

\begin{figure}
\centerline{\psfig{figure=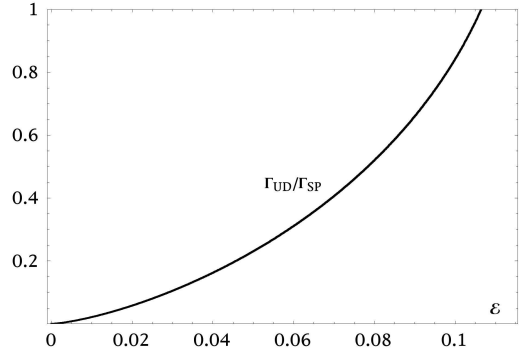,height=4.7in,rheight=3.2in}}
\caption{ The ratio between the Hawking decay rate and the spontaneous emission rate.
For the best confined packet $q=0.9562$ the ratio up to the $\gamma \to 1$ factor is universal for all $n$ and equal $\gamma^4 \Gamma_{UD} / \Gamma_{SP}= 0.190686$ is much smaller but significant and measurable correction.} 
\end{figure}

\bigskip

\centerline{5. CONCLUSIONS}

\medskip
As quantum detectors accelerate they heat up due to the vacuum fluctuations
of the quantum electromagnetic field. No detectors of classical mass can reach the accelerations necessary to observe the effect. 
Trojan atoms as the smallest cyclotrons ever predicted theoretically seem 
to be the best experimental systems to observe the effect.
We show that the Unruh-Davies contribution to the decay of Trojan states 
is as significant as spontaneous emission and therefore readily 
observable effect as equivalent to the Larmor radiation of 
cyclotronic  electrons in the classical limit.
\bigskip

\centerline{6. ACKNOWLEDGEMENTS}
\medskip

We thank Iwo Bialynicki-Birula for valuable comments about electrodynamics
in rotating frames and Kenneth Schafer from providing partial
computer support for initial stages of this research.

\bigskip



\end{twocolumn}


\begin{thebibliography}{000}
\bibitem{expl} The exact general expressions are complicated and contain 
normal trigonometric factors of both types but they greatly simplify in almost static detector limit $\omega \tau \ll 1$ where the resonant factors are dominant.
\bibitem{note1} Nondivergent results can be obtained using expansion 
of Trojan states in hydrogenic basis as Stark states for paramagnetic
Kepler problem \cite{ebe}.
\end{thebibliography}
\end{document}